%% file: 00main.tex
\documentclass[sigconf]{acmart}

\usepackage{soul}
\usepackage{listings}

\usepackage{multirow}
\usepackage{subfigure}
\usepackage{xspace}
\usepackage{rotating}
\usepackage{algorithm}

\usepackage{algorithmic}
\usepackage{mathtools}

\usepackage{amsmath}

\newcommand{\etal}{et~al.\@\xspace}

\newcommand{\financialcnt}{12\xspace}

\newcommand{\platform}{\textit{TermLens}\xspace}
\newcommand{\termname}{unfavorable financial\xspace}
\newcommand{\TermName}{Unfavorable Financial\xspace}
\newcommand{\Termname}{Unfavorable financial\xspace}
\newcommand{\websitecnt}{8,979\xspace}
\newcommand{\termcnt}{1.9 million\xspace}
\newcommand{\websitepct}{42.06\%\xspace}
\newcommand{\termpct}{0.5\%\xspace}
\newcommand{\termtypecnt}{22\xspace}
\newcommand{\termcatcnt}{4\xspace}
\newcommand{\myparagraph}[1]{\textbf{\textit{#1}:}\hspace{3pt}}

\usepackage{pifont} 
\newcommand{\filledCircle}{\ding{108}} 
\newcommand{\halfFilledCircle}{\ding{109}} 

\copyrightyear{2025}
\acmYear{2025}
\setcopyright{cc}
\setcctype{by}
\acmConference[WWW '25]{Proceedings of the ACM Web Conference 2025}{April 28-May 2, 2025}{Sydney, NSW, Australia}
\acmBooktitle{Proceedings of the ACM Web Conference 2025 (WWW '25), April 28-May 2, 2025, Sydney, NSW, Australia}
\acmDOI{10.1145/3696410.3714573}
\acmISBN{979-8-4007-1274-6/25/04}

\begin{document}

\title[Harmful Terms and Where to Find Them]{Harmful Terms and Where to Find Them: Measuring and Modeling Unfavorable Financial Terms and Conditions in Shopping Websites at Scale}

\author{Elisa Tsai}
\affiliation{%
  \institution{University of Michigan}
  \city{Ann Arbor}
  \state{Michigan}
  \country{USA}
}
\email{eltsai@umich.edu}

\author{Neal Mangaokar}
\affiliation{
  \institution{University of Michigan}
  \city{Ann Arbor}
  \state{Michigan}
  \country{USA}
}
\email{nealmgkr@umich.edu}

\author{Boyuan Zheng}
\affiliation{
  \institution{University of Michigan}
  \city{Ann Arbor}
  \state{Michigan}
  \country{USA}
}
\email{boyuann@umich.edu}

\author{Haizhong Zheng}
\affiliation{
  \institution{University of Michigan}
  \city{Ann Arbor}
  \state{Michigan}
  \country{USA}
}
\email{hzzheng@umich.edu}

\author{Atul Prakash}
\affiliation{
  \institution{University of Michigan}
  \city{Ann Arbor}
  \state{Michigan}
  \country{USA}
}
\email{aprakash@umich.edu}


\begin{abstract}
Terms and conditions for online shopping websites often contain terms that can have significant financial consequences for customers. 
Despite their impact, there is currently no comprehensive understanding of the types and potential risks associated with unfavorable financial terms. Furthermore, there are no publicly available detection systems or datasets to systematically identify or mitigate these terms.
In this paper, we take the first steps toward solving this problem with three key contributions.

\textit{First}, we introduce \textit{TermMiner}, an automated data collection and topic modeling pipeline to understand the landscape of unfavorable financial terms.
\textit{Second}, we create \textit{ShopTC-100K}, a dataset of terms and conditions from shopping websites in the Tranco top 100K list, comprising 1.8 million terms from 8,251 websites. Consequently, we develop a taxonomy of 22 types from 4 categories of unfavorable financial terms---spanning purchase, post-purchase, account termination, and legal aspects.
\textit{Third}, we build \textit{TermLens}, an automated detector that uses Large Language Models (LLMs) to identify unfavorable financial terms. 

Fine-tuned on an annotated dataset, \textit{TermLens} achieves an F1 score of 94.6\% and a false positive rate of 2.3\% using GPT-4o. 
When applied to shopping websites from the Tranco top 100K, we find that 42.06\% of these sites contain at least one unfavorable financial term, with such terms being more prevalent on less popular websites. Case studies further highlight the financial risks and customer dissatisfaction associated with unfavorable financial terms, as well as the limitations of existing ecosystem defenses.

\end{abstract}



\maketitle








\input{01introduction}

\input{02related_work}

\input{03data}

\input{04system}

\input{05evaluation}
\input{06limitation}

\input{07conclusion}

\bibliographystyle{ACM-Reference-Format}
\input{00main.bbl}

\input{appendix}

\end{document}

%% file: 01introduction.tex
\section{Introduction}

In 2024, U.S. e-commerce sales are projected to reach approximately \$1.2 trillion~\citep{digitalcommerce2023ecommerce}, with users frequently engaging with websites that impose terms and conditions on financial transactions. While these terms are often benign, they can also facilitate scams or impose unfair consequences on unsuspecting users. This risk is heightened as most users rarely read these lengthy, jargon-filled terms~\citep{obar2020biggest, steinfeld2016agree, bakos2014does}, and are often not required to do so before completing a purchase.

In this work, we define unfavorable financial terms as those that are one-sided, imbalanced, unfair, or malicious, thereby disadvantaging users.~\autoref{fig:example} shows a real-world example of harmful financial terms on a website selling earbuds at seemingly attractive prices. When users make a purchase, the T\&Cs obligate the users to a fitness app subscription with a recurring \$86 monthly fee. \textit{This obligation is not disclosed at all during the purchase process}. 

Unfair financial terms can also exist on legitimate websites---unlike traditional social engineering scams, these terms may not be inherently deceptive but can still cause substantial losses.~\autoref{fig:example2} in Appendix~\ref{sec:case_studies} presents the T\&Cs from Celsius~\citep{celsiuswebsite}, a cryptocurrency company bankrupt in 2022. These terms stipulate that if Celsius goes bankrupt, users could lose digital investments since they would be treated as unsecured creditors. A judge later ruled that Celsius owned its users' cryptocurrency based on these terms~\citep{celsius}, highlighting the real financial risks such terms pose to users. 

It is worth noting that the website in~\autoref{fig:example} operated for at least a year without being flagged by major browsers before its shutdown in June 2024, showing the current defense ecosystem's lack of understanding and mitigation strategies for such \termname terms. Likewise, Celsius's unfair terms only gained attention during bankruptcy proceedings. Despite their impact, few mitigation methods exist for \termname terms.

A possible approach to addressing the concern is to extend the current methods for detecting social engineering scams and dark patterns~\citep{mathur2019dark, mathur2021makes, di2020ui, bongard2021definitely, li2023double, bitaab2023beyond, yang2023trident}. Unfortunately, such an extension is not straightforward. Many of these methods are not designed to detect \termname terms, as they typically focus on content-based features like word patterns, images, website structures, or external indicators like link length and certificates~\citep{bitaab2020scam, bilge2011exposure, kharraz2018surveylance, zhang2011textual, sakurai2020discovering, drury2019certified}, which are unrelated to the detection of \termname terms. Similarly, work on online agreement analysis focuses on privacy policies~\citep{xiang2023policychecker, harkous2018polisis, zhou2023policycomp, bui2023detection} or on terms deemed invalid under the EU law~\citep{lagioia2017automated, braun2021nlp, lippi2019claudette, limsopatham2021effectively, galassi2024unfair, galassi2020cross}.

The emergence of large language models (LLMs) offers a powerful tool for tackling real-world security challenges~\citep{van2022logomotive, li2018fake, zhang2014you, zheng2017smoke, sahingoz2019machine, bitaab2023beyond}, 
 including the analysis of policies like T\&Cs~\citep{rodriguez2024large, frasheri2024llm}, 
but their potential in this area has yet to be fully explored. This paper aims to fill this gap through a large-scale measurement study. To the best of our knowledge, this is the first systematic effort to categorize and detect \termname terms in real-world shopping websites. Our contributions are as follows:

\begin{itemize} 
    \item \myparagraph{Data collection and topic modeling pipeline} We present \textit{TermMiner}\footnote{\url{https://github.com/eltsai/term_miner}}, a scalable pipeline for collecting and analyzing terms and conditions from shopping websites. The pipeline comprises: (1) a T\&Cs collection module, (2) an LLM-based term classification module, and (3) an interactive topic modeling module. \Termname terms are identified using FTC's definitions of unfair practices~\citep{ftcact}.
    
    \item \myparagraph{ShopTC-100K dataset}: We create the \textit{ShopTC-100K} dataset\footnote{\url{https://huggingface.co/datasets/eltsai/ShopTC-100K}}, containing 1.8 million terms extracted from the terms and conditions of 8,251 shopping websites in the Tranco top 100K most popular websites.
    
    \item \myparagraph{\Termname term taxonomy}: We develop a comprehensive taxonomy for \termname terms, covering \termcatcnt categories and \termtypecnt types, including terms related to purchase, billing, post-purchase activities, account termination and recovery, and legal conditions.
    
    \item \myparagraph{\Termname terms detection}: We develop \textit{Term-Lens}, an LLM-based detection framework using LLMs to automatically identify \termname terms. With a fine-tuned GPT-4o model, \platform achieves a 94.6\% F1 score and a 2.3\% FPR on an annotated evaluation dataset. 
    
    \item \myparagraph{Measurement study}: We analyze \termcnt terms from \websitecnt online shopping websites, finding that \websitepct of English websites in the Tranco top 100k contain \termname terms. Our analysis reveals these terms are more common on less popular websites, with case studies highlighting the potential financial and legal harm to consumers. 
\end{itemize}

\input{fig_tex/scam_example}

%% file: fig_tex/scam_example.tex
\begin{figure}[!t] 
 \centering
 \includegraphics[width=0.99\columnwidth]{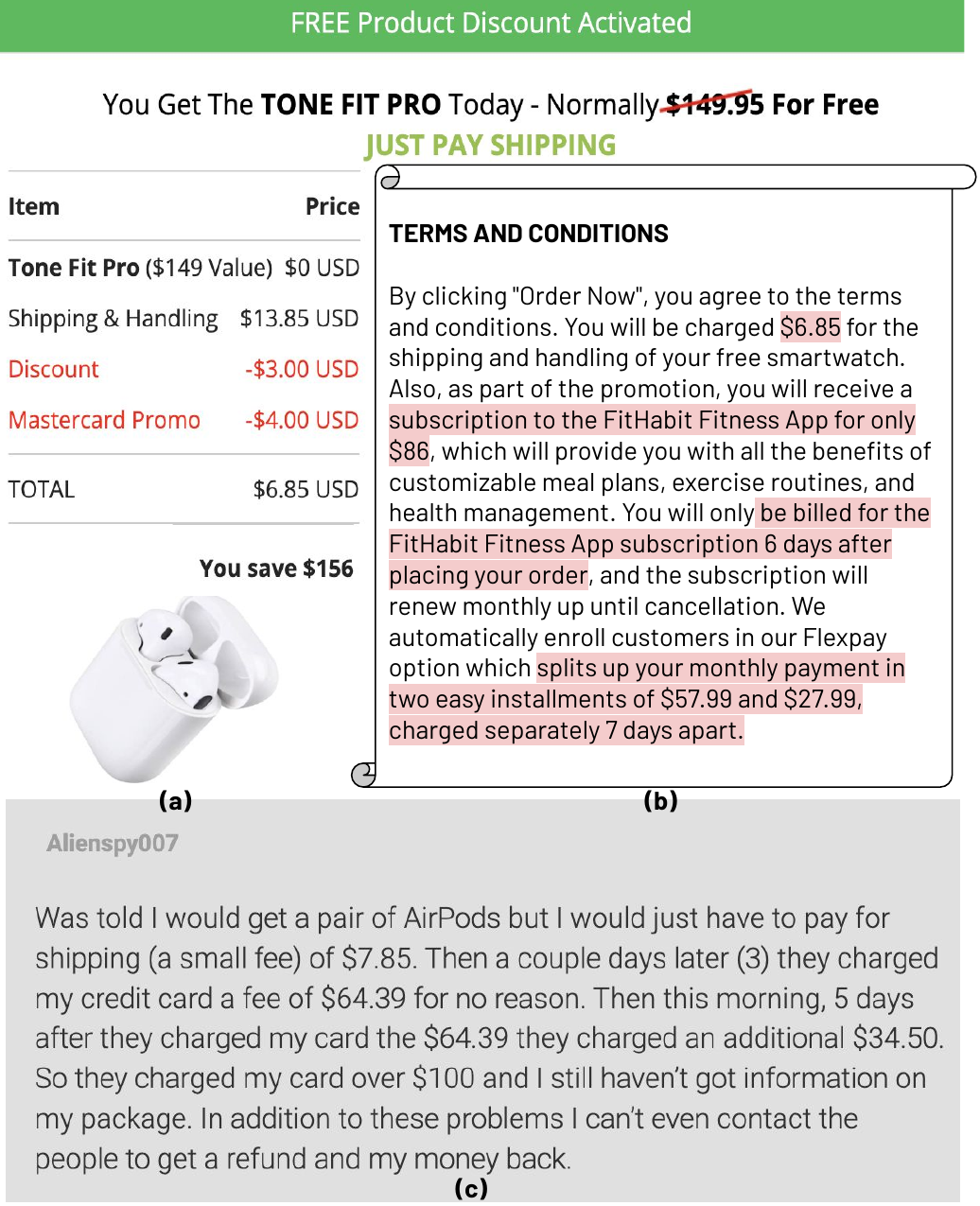}
 \caption{\textbf{\Termname term example} --- (a) shows the payment page for Tone Fit Pro, a now-defunct website, with no mention of the subscription service on the payment page. (b) displays its T\&Cs, stating customers are automatically enrolled in an \$86/month Fitness App subscription with auto-renewal. (c) shows a screenshot of real-life victim complaints.}
 \Description[Scam website example with hidden auto-renewal term.]
 {This figure illustrates an example of a deceptive financial term found in a scam website. 
 (a) The payment page for "Tone Fit Pro" lacks any mention of a subscription service.
 (b) The terms and conditions reveal that users are automatically enrolled in an \$86/month FitHabit Fitness App subscription with auto-renewal.
 (c) A compilation of real-world consumer complaints highlights user frustration and confusion over unexpected charges.}
\label{fig:example}
\end{figure}

%% file: 02related_work.tex
\section{Related Work}

\myparagraph{Scam and fake e-commerce website detection}
Detection methods for scam and fake e-commerce websites (FCW) typically rely on two types of features: external (e.g., URLs, certificates, logos, redirect mechanisms)~\citep{blum2010lexical, zouina2017novel, moghimi2016new, sakurai2020discovering, drury2019certified, van2022logomotive, li2018fake, zhang2014you, zheng2017smoke, sahingoz2019machine, bitaab2023beyond} and content-based (e.g., visual and HTML structures, images, scripts, hyperlinks)~\citep{xiang2011cantina+, kharraz2018surveylance, yang2019phishing, jain2017phishing, bitaab2023beyond, yang2023trident}. These models are either rule-based or machine learning-based, with feature selection grounded in domain knowledge (e.g., indicative images, third-party scripts). However, no prior work in this line has considered terms and conditions and their financial impacts on users.

We consider social engineering scams to overlap with our detection target. The \termname terms in \autoref{fig:example} function similarly by deceiving users into signing up for additional subscriptions. However, as discussed in \S\ref{sec:financial_terms_section} and~\S\ref{sec:categoring}, \termname terms are not exclusive to scam websites. Therefore, consumers should be alerted to the presence of such terms. We view our work as the first to measure \termname terms at scale.

\input{fig_tex/data_collection_pipeline}

\myparagraph{Dark patterns}
Dark patterns are deceptive user interface designs intended to manipulate users into actions against their best interests~\citep{mathur2019dark}. Recent research has examined their psychological impact on user decision-making~\citep{mathur2021makes,nouwens2020dark,waldman2020cognitive,narayanan2020dark}, while also exploring legal frameworks and strategies for intervention~\citep{luguri2021shining,gray2021dark}.

Although terms and conditions are not part of the user interface design, we consider the \termname terms we identify to be closely related to dark patterns. The unilateral nature of these terms and their potential to hide uncommon or unexpected terms make them closely align with the characteristics of dark patterns: asymmetric, covert, deceptive, hiding information, and restrictive.

\myparagraph{Terms and conditions legal analysis} 
There is limited NLP-based analysis of legal documents like online contracts and terms of service~\citep{lagioia2017automated, braun2021nlp, lippi2019claudette, limsopatham2021effectively, jablonowska2021assessing, galassi2024unfair}. Prior studies, such as Lippi \etal~\citep{lippi2019claudette} and Galassi \etal~\citep{galassi2024unfair}, typically focus on small datasets of T\&Cs (25 to 200 documents). However, their focus is mainly on assessing fairness under the EU’s Unfair Contract Terms Directive~\citep{CouncilDirective1993} (i.e., clauses invalid in court). In contrast, our work specifically targets terms with more direct financial impacts on users.

In this paper, we focus on the financial terms in the large-scale measurement of terms and conditions from English shopping websites, assessed using the definition of unfair acts or practices as provided by the Federal Trade Commission (FTC)'s Policy Statement on Deception~\citep{ftc1983deception}. A detailed comparison of our proposed term taxonomy with prior work is provided in Appendix~\ref{sec:appendix_other_templates}.

\myparagraph{Privacy policy analysis}
A significant body of work investigates the viability of NLP-based analysis for privacy policies. One significant line of such research focuses on detecting contradicting policy statements (e.g., via ontologies~\citep{andow2019policylint} and knowledge graphs~\citep{cui2023poligraph}) or ambiguities~\citep{shvartzshnaider2019going}. Other areas include improving user comprehension~\citep{harkous2018polisis}, topic modeling, and summarization~\citep{alabduljabbar2021automated, sarne2019unsupervised}.

In this work, we focus on financial terms which are distinct from privacy policies. While we also perform topic modeling, we are the first to apply such a pipeline to construct a taxonomy for \termname terms. Furthermore, detecting contradictions and ambiguities is orthogonal to the detection of malicious financial terms, making it difficult to apply similar techniques directly.

%% file: fig_tex/data_collection_pipeline.tex
\begin{figure*}[t] 
 \centering
 \includegraphics[width=2\columnwidth]{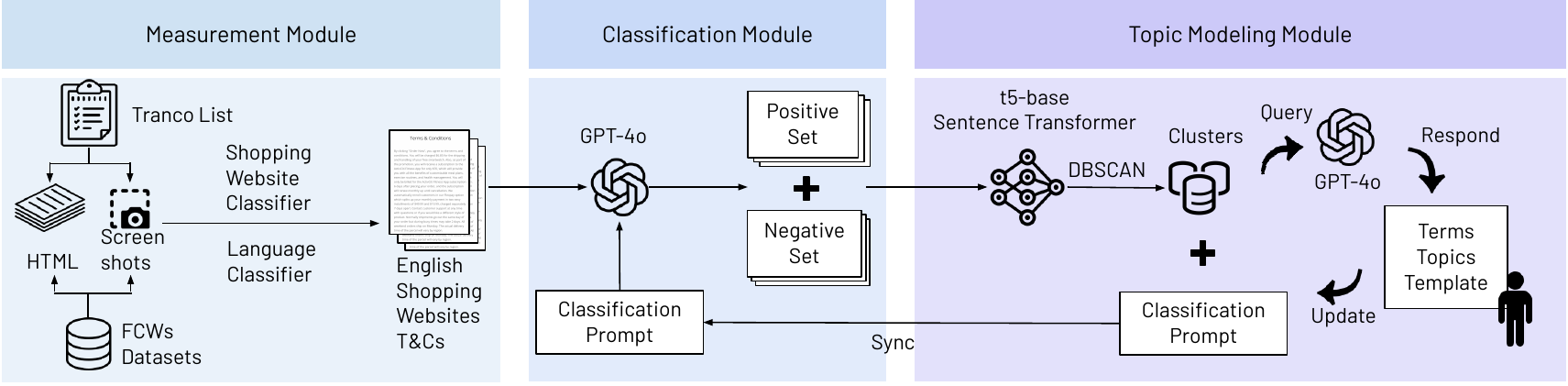}
 \caption{\textbf{\textit{TermMiner} (data collection and topic modeling pipeline)}---(1) Measurement module: collects shopping websites from the Tranco list and fake e-commerce website datasets, extracting English terms and conditions from shopping websites. (2) Term classification module: classifies the terms into binary categories based on a given prompt. (3) Topic modeling module: leverages t5-base Sentence Transformer and DBSCAN for clustering. Topics are derived from the clusters using a combination of manual inspection and GPT-4o, employing a snowball sampling method~\citep{goodman1961snowball} to iteratively develop a topic template of terms.}
 \Description[Pipeline for collecting and analyzing unfavorable financial terms.]
 {This figure presents the \textit{TermMiner} pipeline for collecting and analyzing unfavorable financial terms from shopping websites. 
 (1) The measurement module gathers websites from the Tranco list and known fake e-commerce datasets, extracting English terms and conditions from shopping sites.
 (2) The term classification module processes extracted terms, categorizing them into binary labels based on a predefined prompt.
 (3) The topic modeling module applies the T5-base Sentence Transformer and DBSCAN clustering to group terms into topics. These topics are refined through a combination of manual inspection and GPT-4o, using a snowball sampling approach to iteratively develop a comprehensive topic taxonomy.}
\label{fig:financial_term_pipeline}
\end{figure*}

%% file: 03data.tex
\section{Understanding Unfavorable Financial Terms}
\label{sec:topic_modeling_section}

In this section, we outline our detection goal and present \textit{TermMiner}, a pipeline for collecting, clustering, and topic modeling \termname terms from English shopping websites in the Tranco top 100,000~\citep{tranco} and datasets of fraudulent e-commerce sites~\citep{bitaab2023beyond, janaviciute2023fraudulent}. As shown in \autoref{fig:financial_term_pipeline}, the pipeline categorizes two types of terms: (1) financial terms that may have immediate or future financial impacts, and (2) \termname terms, identified as unfair, unfavorable, or concerning for customers. We then summarize the taxonomy of \termname terms, which fall into four broad categories: (1) purchase and billing, (2) post-purchase, (3) termination and account recovery, and (4) legal terms.

\subsection{Threat Model}
\label{ssec:ftc}

We aim to detect one-sided, imbalanced, unfair, or malicious \textit{financial} terms in online shopping websites' terms and conditions, which could pose significant risks to users, potentially leading to unexpected financial losses. These risks can arise from website operators seeking to limit liability or from intentional malfeasance.

To assess whether a financial term is unfavorable, we refer to Section 5 of the Federal Trade Commission (FTC) Act~\citep{ftcact}, which defines an act as unfair if it meets the following criteria:

\begin{itemize}
    \item \textbf{C1: Substantial Injury}. It causes or is likely to cause substantial injury to consumers;
    \item \textbf{C2: Unavoidable Harm}. Consumers cannot reasonably avoid it; and
    \item \textbf{C3: Insufficient Benefits}. It is not outweighed by countervailing benefits to consumers or competition. 
\end{itemize}

During the topic modeling of term clusters, we judge the topic representing each cluster by three criteria to evaluate their fairness.

\myparagraph{C1} Since we focus on financial terms with potentially detrimental impacts, all financial terms inherently satisfy this criterion.

\myparagraph{C2} Terms and conditions are often hidden or difficult to avoid. Fair financial terms must be clearly displayed at critical points, like the payment page. However, terms related to cancellation, refunds, and returns are rarely shown upfront. We evaluate terms for unexpected fees (e.g., cancellation charges, non-refundable items, costly returns) that place an undue burden on consumers.

\myparagraph{C3} We classify terms as benign if they serve legitimate user or business protection, such as terms prohibiting fraud or abuse, protecting intellectual property, or ensuring legal compliance.

\subsection{Data Collection and Topic Modeling}
\label{ssec:data_collection_and_topic_modeling}

As shown in~\autoref{fig:financial_term_pipeline}, we introduce \textit{TermMiner}, a data collection and topic modeling pipeline for identifying \termname term at scale.  By integrating LLMs like GPT-4o~\citep{openai2023gpt4}, \textit{TermMiner} significantly reduces the extensive manual efforts required in previous web content mining studies, such as those focused on detecting dark patterns~\citep{mathur2019dark}. 
\textit{TermMiner} is open-sourced and can be repurposed for various web-based text analysis tasks or longitudinal studies. Researchers can use our tools to explore different aspects of terms and conditions, such as readability, accessibility, or fairness.

\myparagraph{A Two-Pass Method}
In the data collection and topic modeling steps, we employ a two-pass method. The first pass focuses on modeling and detecting \textit{financial terms} to develop a corresponding topic template. In the second pass, we use the detected financial terms to re-conduct the classification and topic modeling modules. This time, the goal is to detect \termname terms within the financial terms identified. This approach is necessary because, to the best of our knowledge, there are no established templates or annotation schemes for (1) financial terms or (2) \termname terms in online shopping agreements. This two-pass process ensures comprehensive detection and accurate categorization of both financial and \termname terms.

\myparagraph{(1) Measurement Module} The measurement module collects terms and conditions from shopping websites to build a large, diverse dataset for analysis. For our large-scale measurement, we collect English shopping websites from two sources: the Tranco list~\citep{tranco}, a ranking of top global websites, and two fraudulent e-commerce datasets (FCWs~\citep{bitaab2023beyond} and the Fraudulent and Legitimate Online Shops Dataset~\citep{janaviciute2023fraudulent}). We filter out non-English content using Python's langdetect library~\citep{langdetect}. To classify shopping websites, we evaluate several configurations: (1) GPT-3.5-Turbo~\citep{gpt35} with URL, (2) GPT-3.5-Turbo with URL and HTML content, (3) GPT-4o~\citep{openai2023gpt4} with URL, and (4) GPT-4o with URL and website screenshot. To evaluate our classification methods, we manually annotated a sample of 500 websites from the Tranco list, categorizing them into ``shopping'' and ``non-shopping.'' GPT-4o, when prompted with URLs and screenshots, achieved an accuracy of 92\%, comparable to commercial website classification services~\citep{mathur2019dark} (see Appendix~\ref{sec:website_cls} for details). Therefore, we use this configuration throughout our work.

We subsequently crawl the shopping websites to collect terms and conditions pages. A snowballed regex matching method detects terms and any nested policy pages, refined through positive and negative regex patterns to improve accuracy. Starting with common anchor texts, we iteratively refine the regex patterns by analyzing T\&C links, which can be found in Appendix~\ref{sec:appendix_reg}. As shown in~\autoref{table:dataset_stats}, we collected \termcnt terms from \websitecnt websites in total.

\myparagraph{(2) Classfication Module}
The classification module categorizes terms from shopping websites' terms and conditions into binary categories: positive or negative. The categorization is based on the detection goal (such as identifying financial terms or identifying \termname terms) using corresponding prompts with the GPT-4o model~\citep{openai2023gpt4}.

We opt for prompt engineering instead of fine-tuning the LLM for term classification to reduce costs. Prior work~\citep{sun2023text,openai2023bestpractices}, along with our empirical observation (see~\S\ref{sec:eva}), indicates that clear task descriptions and relevant examples (taxonomy) significantly enhance LLM performance in text classification. For \termname terms, we use the ``Unfavorable Term Taxonomy Prompt'' from Appendix~\ref{sec:prompts} and topics identified in the topic modeling step, to perform zero-shot term classification on a given set of terms and conditions. This process outputs sets of positive and negative terms, which are then used for clustering, inspection, and topic modeling. The resulting template generated from this analysis will, in turn, enhance the classification accuracy, creating a feedback loop that continuously improves our detection capabilities. 

\input{tables/dataset_stats}

\myparagraph{(3) Topic Modeling Module} The topic modeling module uses LLMs and manual inspection to organize terms into meaningful topics. We generate sentence embeddings with the T5 model~\citep{raffel2020exploring} and apply the DBSCAN clustering algorithm~\citep{ester1996density} to group terms by semantic similarity. The DBSCAN hyperparameters are decided through manual inspection. To extract high-frequency topics, we leverage GPT-4o~\citep{openai2023gpt4}, building on recent findings that show LLMs outperform traditional topic modeling methods like Latent Dirichlet Allocation (LDA)~\citep{blei2003latent} and BERTopic~\citep{grootendorst2022bertopic} in topic analysis~\citep{shrestha2023we, mu2024large}.

We develop an iterative topic modeling approach assisted by GPT-4o proceeds as follows: 
\begin{enumerate}
    \item We analyze DBSCAN clusters and create an initial topic template for financial terms.
    \item GPT-4o performs topic modeling on random samples from each cluster, assigning them to existing topics or suggesting new ones.
    \item We review and refine new topic suggestions through manual inspection, and updating the template.
    \item This process iterates until all clusters are assigned to a meaningful and satisfactory topic.
\end{enumerate}

This iterative workflow, combining clustering, human-guided template creation, and GPT-4o's advanced topic modeling capacity, enables efficient and comprehensive extraction of the topic template. We analyze 22,112 clusters in total, creating the \termname term taxonomy below.

\input{tables/malicious_terms}

\myparagraph{ShopTC-100K Dataset.} In the data collection stage, we extract 8,251 shopping websites from the Tranco top 100K, yielding 1.8 million terms.~\autoref{tab:dataset_stats} presents ShopTC-100K statistics alongside two fake e-commerce datasets, with unfavorable financial terms identified in later measurement study (\S\ref{sec:findings}).

\subsection{\TermName Term Taxonomy}
\label{sec:financial_terms_section}

We categorize \termtypecnt types of \termname terms into \termcatcnt categories (\autoref{table:taxonomy}) and provide real-world examples analyzed against the three
criteria proposed by the FTC Act criteria (\S\ref{ssec:ftc})~\citep{ftcact}. A detailed taxonomy is in Appendix~\ref{sec:detailed_tax}. While not inherently deceptive, these terms often impose financial obligations consumers should recognize. We note that the severity of such terms depends on \textit{context}, which we leave for future research. We do not claim this list is exhaustive; however, it represents the most prominent types among the \termcnt terms from \websitecnt websites. We also report the financial term template in Appendix~\ref{sec:appendix_finaincial_terms}.

It is important to note that this paper \textit{does not} aim to analyze the fairness of terms from the legal perspective. We consider our work to be a complementary addition to the AI \& Law datasets~\citep{lippi2019claudette, galassi2024unfair}, by focusing on the natural phrasing found in online shopping websites' terms and conditions. A comparison between our \termname term template and previous work on online agreement fairness can be found in Appendix~\ref{sec:appendix_other_templates}.


%% file: tables/dataset_stats.tex
\begin{table}[!t]
\centering
\footnotesize
\caption{Dataset and detection statistics---We source data from the Tranco top 100k list, the FCWs dataset~\citep{bitaab2023beyond}, and the FLOS (Fraudulent and Legitimate Online Shops Dataset)~\citep{janaviciute2023fraudulent}, resulting in a total of \websitecnt English shopping websites with terms and conditions. We report the statistics of detected \termname terms within them. }
\label{table:dataset_stats}

\begin{tabular}{p{0.1cm}p{1.8cm}p{1.3cm}p{0.8cm}p{0.8cm} |p{1.1cm}}
\toprule
& \textbf{Source} & \textbf{ShopTC-100K (Ours)} & \textbf{FCW} & \textbf{FLOS} & \textbf{Total} \\
\midrule
\multirow{5}{*}{\rotatebox[origin=c]{90}{Datasets}} & Website to Query 
& 100,000 
& 6,127 
& 1,040 
& 107,167 \\
\cmidrule(lr){2-6}

& Shopping 
& 61,466 
& 1,378
& 542 
& 63,386 \\
\cmidrule(lr){2-6}

& English 
& 38,674 
& 1,157
& 317 
& 40,148 \\

\cmidrule(lr){2-6}
& \shortstack[l]{Website with T\&C} 
& 8,251 
& 463 
& 265 
& 8,979 \\
\cmidrule(lr){2-6}

& Term Count 
& 1,825,231 
& 56,921
& 27,604
& 1,909,756 \\

\midrule

& \shortstack[l]{Website with \\Unfavorable Terms} 
&  3,633
&  175
&  165
&  3,973 \\

\midrule

\multirow{6}{*}{\rotatebox[origin=c]{90}{Number of Detected Terms}}
& \shortstack[l]{Purchase and\\Billing} 
&  2,323
&  62
&  41
&  2,426 \\

\cmidrule(lr){2-6}
& \shortstack[l]{Post-Purchase} 
& 6,143
& 378
& 363
& 6,884
\\

\cmidrule(lr){2-6}
& \shortstack[l]{Termination and\\Account Recovery} 
& 220
& 2
& 3
& 225
\\

\cmidrule(lr){2-6}
& Legal 
& 558
& 14
& 3
& 575 \\
\cmidrule(lr){2-6}

& Others 
& 40
& 0
& 0
& 40 \\

\cmidrule(lr){2-6}
& \shortstack[l]{Total Unfavorable\\Financial Terms}    
& 9,284
& 456
& 410
& 1,0150 \\
\bottomrule
\end{tabular}
\end{table}

%% file: tables/malicious_terms.tex
\begin{table*}[t!]
    \footnotesize
    \centering
    \caption{{Categories, types, and examples of \termname terms are clustered, extracted, and topic-modeled from \termcnt terms across \websitecnt websites. All examples are extracted as-is from real-world shopping websites. The criteria are as follows: C1 = ``Substantial Injury'' (the term causes or is likely to cause substantial injury to consumers), C2 = ``Unavoidable Harm'' (consumers cannot reasonably avoid it), C3 = ``Insufficient Benefits'' (it is not outweighed by countervailing benefits to consumers or competition). The symbols represent the likelihood of satisfaction of a given criterion: \filledCircle = Always, \halfFilledCircle = Sometimes.}}
\begin{tabular*}{1.96\columnwidth}{p{1cm} p{4cm} p{9cm}  p{0.2cm} p{0.2cm} p{0.2cm}}
    \toprule
    \textbf{Category} & \textbf{Type} & \textbf{Example} & C1 & C2 & C3 \\
    \midrule

    \multirow{11}{*}{\shortstack{Purchase\\and\\Billing\\Terms}}

     & \shortstack{Immediate Automatic Subscription} 
    & {Also, as part of the \hl{promotion}, you will receive a \hl{subscription to the FitHabit Fitness App for only \$86}, and the subscription will renew monthly up until cancellation.} 
    & \filledCircle & \halfFilledCircle & \filledCircle \\
    \cmidrule(lr){2-6}
    & {Automatic Subscription after Free Trial}
    & {After the Promotion period has ended, unless you cancel the service before the end of the free trial period, you will \hl{automatically be subscribed} onto the regular paid 1-year plan at the price of \$275.40, which will automatically renew for successive 12-month periods, until cancelled.}
     & \filledCircle &  \halfFilledCircle & \halfFilledCircle\\
    \cmidrule(lr){2-6}
    & {Unilateral Unauthorized Account Upgrades} 
    & {Brevo reserves the right to automatically increase the contacts limit in the User account and \hl{upgrade the User’s plan without prior notice}. } 
     & \filledCircle & \filledCircle & \filledCircle \\
    \cmidrule(lr){2-6}
    & {Late or Unsuccessful Payment Penalty}
    & {In addition, if any payment is not received within \hl{30 days after the due date}, then we may charge a \hl{late fee} of \$10 and we may assess interest at the rate of 1.5\% of the outstanding balance per month (18\% per year), or the maximum rate permitted by law.}
     & \filledCircle & \halfFilledCircle & \halfFilledCircle \\
    \cmidrule(lr){2-6}
    & {Overuse Penalty}
    & {If the Company establishes limits on the frequency with which you may access the Site, or terminates your access to or use of the Site, you agree to pay the Company one hundred dollars (\$100) for each message posted \hl{in excess of such limits} or for each day on which you access the Site in excess of such limits, whichever is higher.}
     & \filledCircle & \halfFilledCircle & \halfFilledCircle \\
    \cmidrule(lr){2-6}
    & {Retroactive Application of Price Change} 
    & {When an applicable exchange rate is updated or when a change of price is notified to Brevo by its suppliers or WhatsApp, Brevo might \hl{immediately apply with retroactive effect} the new Ratio and price increase to the User.} 
     & \filledCircle & \filledCircle & \halfFilledCircle \\

    \midrule
    \multirow{11}{*}{\shortstack{Post-\\Purchase\\Terms}}
    & {Non-Refundable Subscription Fee}
    & {If you or we cancel your subscription, you are \hl{not entitled to a refund of any subscription fees} that were already charged for a subscription period that has already begun.}
     & \halfFilledCircle & \halfFilledCircle & \filledCircle \\
    \cmidrule(lr){2-6}
    & {No Refund For Purchase} 
    & {Unless a refund is required by law, there are \hl{No Refund For Purchases for POS terminals} and all transactions are final.}
     & \halfFilledCircle & \halfFilledCircle &  \halfFilledCircle \\
     
    \cmidrule(lr){2-6}
    & {Strict No Cancellation Policy}
    & {As Research and Markets starts processing your order once it is submitted, we operate a \hl{strict no cancellation policy}.}
     & \halfFilledCircle & \halfFilledCircle &  \halfFilledCircle \\

    \cmidrule(lr){2-6}
    & {\shortstack{Cancellation Fee or Penalty}}
    & {Some Bookings \hl{can’t be canceled for free}, while others can only be canceled for free before a deadline.}
     & \halfFilledCircle & \halfFilledCircle & \halfFilledCircle\\
    \cmidrule(lr){2-6}
    & Non-Refundable Additional Fee
    & {For this service, National Park Reservations charges a \hl{10\% non-refundable reservation fee} based on the total dollar amount of reservations made. }
     & \halfFilledCircle & \halfFilledCircle & \halfFilledCircle\\

    \cmidrule(lr){2-6}
    & {Non-Monetary Refund Alternatives} 
    & {Refund Policy: Refunds are not in cash but in the form of a \hl{``coupon’’}.}
    & \halfFilledCircle & \halfFilledCircle & \filledCircle \\
    \cmidrule(lr){2-6}
    
    & No Responsibility for Delivery Delays 
    & {We will \hl{not be held responsible} if there are \hl{delays in delivery} due to out-of-stock products. } 
    & \halfFilledCircle & \halfFilledCircle & \filledCircle \\
    \cmidrule(lr){2-6}
    & {Customers Responsible for Shipping Issues} 
    &  {If the parcel is on hold by the Customs department of the shipping country, \hl{the customer is liable} to provide all relevant and required documentation on to the authorities. Asim Jofa is \hl{not liable to refund} the amount in case of \hl{non-clearance of the parcel}.} 
    & \halfFilledCircle & \halfFilledCircle & \filledCircle\\
    \cmidrule(lr){2-6}
    & Customers Pay Return Shipping 
    & {All shipping costs will have to be borne by the customer.} & \halfFilledCircle & \halfFilledCircle & \filledCircle \\
    \cmidrule(lr){2-6}
    & {\shortstack{Restocking Fee}}
    & {An \hl{8\% restocking fee} and shipping fees for both ways will be borne by the buyer if returned without defects within 30 days from the purchase date or 7 days from delivery date, whichever is later.}
    & \halfFilledCircle & \halfFilledCircle & \filledCircle \\

    \midrule
    \multirow{3}{*}{\shortstack{Termination\\and\\Account\\Recovery\\Terms}} 
    & {Account Recovery Fee}
    &  {To recover an archived or locked account, the legitimate creator of the account shall provide verifiable information about one's identity and will be charged a \hl{10\% administrative fee} for the additional work caused by the account recovery process.} 
    & \halfFilledCircle & \halfFilledCircle & \halfFilledCircle\\
    
    \cmidrule(lr){2-6}
    & {Digital Currency, Reward, Money Seizure on Inactivity}
    & {Please be noted that if your account is \hl{dormant} for a period of 12 consecutive calendar months or longer, ..., any amounts in your account’s balance, including any outstanding fees owed to you, shall be considered as \hl{forfeited and shall be fully deducted} to Appnext.}
    & \halfFilledCircle & \halfFilledCircle & \filledCircle \\
    
    \cmidrule(lr){2-6}
    & {Digital Currency, Reward, Money Seizure on Termination or Account Closure} 
    &  {All Currency and/or Virtual Goods shall be \hl{cancelled} if Your account is \hl{terminated} or suspended for any reason or if We discontinue providing the Games and we will not compensate you for this loss or make any refund to you.}
    & \halfFilledCircle & \halfFilledCircle & \halfFilledCircle\\


    \midrule
    \multirow{4}{*}{\shortstack{Legal\\Terms}}
    & Exorbitant Legal Document Request Fee 
    & {Responding to requests for production of documents, and other matters requiring more than mere ministerial activities on our part, will incur a fee of \hl{two hundred dollars (\$200) per hour.}}
    & \halfFilledCircle & \halfFilledCircle & \halfFilledCircle\\
    \cmidrule(lr){2-6} 
    &  Forced Waiver of Legal Protections
    &  {You hereby \hl{waive California Civil Code Section 1542}. You hereby waive any similar provision in law, regulation, or code.}
    & \halfFilledCircle & \halfFilledCircle & \halfFilledCircle\\
    \cmidrule(lr){2-6} 
    & {Forced Waiver of Class Action Rights}
    & {This agreement includes a \hl{class action waiver} and an arbitration provision that governs any disputes between you and Sendinblue.}
    & \halfFilledCircle & \halfFilledCircle & \halfFilledCircle\\
    \cmidrule(lr){2-6} 
    & {Other Legal Unfavorable Financial Term}
    & {...}\\
    
    \bottomrule
\end{tabular*}
\label{table:taxonomy}

\end{table*}

%% file: 04system.tex
\section{\TermName Term Detection System}
\label{sec:detection_section}

\input{fig_tex/plugin_design}

In this section, we introduce \platform, a Chrome plugin designed to detect \termname terms on e-commerce websites. Built upon the insights gained from the \termname term template and topic modeling analysis, \platform enables efficient identification of potentially harmful financial terms, providing users with real-time protection against \termname terms.

\subsection{System Overview}

Our detection system is illustrated in \autoref{fig:plugin}. When a user activates \platform, the URL of the current page is sent to the backend. Upon receipt, the backend crawler collects the terms and conditions pages. These term pages, along with the HTML content of the current page (and a screenshot if paired with a multimodal LLM), are preprocessed and sent to the pluggable LLM module for further analysis. If the LLM module flags any terms as \termname terms, the alert generator sends the identified terms back to the frontend, where they are displayed to the user.

\myparagraph{Pluggable LLM Module} 
We parse the current page to determine if it is a payment page, improving alert accuracy by cross-checking terms with payment page details. For example, in \autoref{fig:example}, the term ``You will be charged \$6.85 for the shipping and handling of your free smartwatch'' aligns with the payment page, making it less concerning than the Immediate Automatic Subscription term, ``you will receive a subscription to the FitHabit Fitness App for only \$86,'' which is not shown on the payment page.

The Pluggable LLM Module, a key part of our system, analyzes both terms and conditions pages and the current webpage. By keeping the LLM decoupled from the backend, we allow flexibility in integrating different models. This enables multimodal models like GPT-4, GPT-4o~\citep{openai2023gpt4}, or LLaMA 3.2~\citep{llama3.2-90B-vision} to process screenshots and terms, or text-based models such as GPT-3.5~\citep{gpt35}, LLaMA~\citep{touvron2023llama}, Mistral~\citep{jiang2023mistral}, or Gemma~\citep{team2024gemma} to analyze HTML and terms.

\input{tables/annotated_dataset}

\myparagraph{Backend Core Module}
The alert generator receives flagged \termname terms and checks if the user is on a payment page. If so, it only flags terms not displayed on that page. GPT-4o analyzes page screenshots to mimic the user's experience and guard against adversarial text-based evasion. When the page is not a payment page, all flagged financial terms are shown. Since returns and refunds are rarely disclosed on payment pages, our evaluation in~\S\ref{sec:eva} focuses on scenarios where the user is not on a payment page and seeks to assess financial risks in advance.

%% file: fig_tex/plugin_design.tex
\begin{figure}[!t] 
 \centering
 \includegraphics[width=0.99\columnwidth]{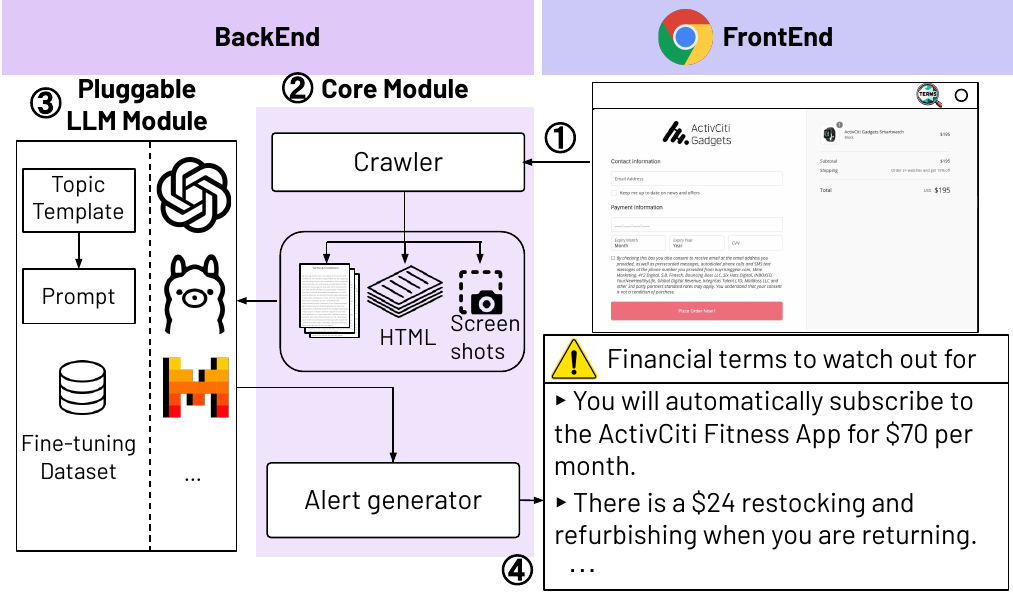}
 \caption{\textbf{\platform Design}---
  (1) When the user activates the plugin, the current page URL is sent to the backend. (2) The terms and conditions are crawled and combined with the page information. (3) The pluggable LLM module analyzes the data to identify unfavorable financial terms. (4) Alerts are generated and displayed on the front end to warn users of potentially unfair financial terms. }
\Description[Diagram of \platform plugin workflow.]
 {This figure illustrates the design of the \platform plugin, which detects unfavorable financial terms on shopping websites. 
 (1) When a user activates the plugin, it sends the current page URL to the backend.
 (2) The backend crawls and extracts the terms and conditions associated with the page.
 (3) A pluggable LLM module processes the extracted data to detect potentially unfair financial terms.
 (4) If any unfavorable terms are identified, the plugin generates an alert and displays it on the front end to warn the user.}
\label{fig:plugin}
\end{figure}

%% file: tables/annotated_dataset.tex
\begin{table}[t!]
    \centering
    \footnotesize
    \caption{Statistics of annotated datasets for fine-tuning and validation for each term category.}
    \label{tab:dataset_stats}
    \begin{tabular}{p{3.5cm} p{1.5cm} p {1.5cm}}
        \toprule
        \textbf{Type} & \textbf{Fine-tuning} & \textbf{Validation} \\
        \midrule

        Post-Purchase & 51 & 48 \\
        Legal & 30 & 30 \\
        Termination and Account Recovery & 15 & 16 \\
        Purchase and Billing & 32 & 32 \\
        \midrule
        Unfavorable Terms Combined & 128 & 126 \\
        \midrule
        Benign & 116 & 119 \\
        \midrule
        Total Count & 244 & 245 \\
        \bottomrule
    \end{tabular}
\end{table}

%% file: 05evaluation.tex
\section{Evaluation and Large-Scale Measurement}

\input{tables/classification_results}

We implement and evaluate \platform using a manually annotated dataset. Our evaluation focuses on two key aspects: (1) assessing detection performance to determine how effectively LLMs, including both zero-shot and fine-tuned models, identify \termname terms (\S\ref{sec:eva}), and (2) analyzing findings from large-scale measurements using \platform (\S\ref{sec:findings}).

\subsection{Evaluation on an Annotated Dataset}
\label{sec:eva}

\myparagraph{Dataset} We create an annotated dataset by randomly selecting 500 terms from clusters of both \termname terms and negative clusters (i.e., benign financial or non-financial terms). This yields 250 potential \termname terms and 250 benign terms. Three researchers independently labeled the terms using the \termname template, without knowledge of the clusters. Disagreements were resolved in a second pass, and duplicates were removed, resulting in 489 final terms. The dataset was split into 244 terms for fine-tuning and 245 terms for validation (\autoref{tab:dataset_stats}).

\myparagraph{Baselines}
To our knowledge, no prior work has directly addressed the detection of unfavorable financial terms. Recent advances in large language models (LLMs) demonstrate superior performance in common sense reasoning, complex text classification, and contextual understanding~\citep{gpt35,openai2023gpt4,touvron2023llama}, outperforming older models like BERT~\citep{devlin2018bert} and RoBERTa~\citep{liu2019roberta}. Therefore, we evaluate state-of-the-art LLMs: (1) GPT-3.5-Turbo~\citep{gpt35}, (2) GPT-4-Turbo~\citep{openai2023gpt4}, and (3) GPT-4o, along with two open-source LLMs: (1) LLaMA 3 8B~\citep{touvron2023llama} and (2) Gemma 2B~\citep{team2024gemma}.

\myparagraph{Evaluation Configurations}
We evaluate two configurations: (1) Zero-shot classification with a simple binary prompt describing the unfavorable financial term and a multi-class taxonomy prompt explaining term types, and (2) Fine-tuning the LLM using the taxonomy to improve detection accuracy (see prompts in Appendix~\ref{sec:prompts}).

\myparagraph{Metrics} We evaluate the models in terms of false positive rate, true positive rate, F1 score, and Area Under the Curve. AUC represents the area under the ROC (Receiver Operating Characteristic) curve, measuring the model's ability to distinguish between classes.


\myparagraph{Zero-shot Classification Performance}
As a baseline for \termname term detection, we evaluated zero-shot classification with two prompts: (1) a simple prompt defining unfavorable financial terms and (2) a taxonomy prompt explaining term types. Using the taxonomy improved the True Positive Rate (TPR) by 4.4\% to 27.4\% and boosted the F1 score by 4.5\% to 21.1\%, showing a better balance of precision and recall. However, the False Positive Rate (FPR) increased in most cases, except for GPT-4o, where it dropped by 24.2\%. GPT-4o achieved the best overall performance with a TPR of 96.6\% and an F1 score of 82.5\%, demonstrating the importance of a \termname term taxonomy for more accurate detection.

\myparagraph{Fine-tuned LLM Classification Performance}
We fine-tune GPT-3.5-Turbo and GPT-4o for 4 epochs with a batch size of 1. Fine-tuning resulted in significant performance improvements, with GPT-4o achieving a True Positive Rate (TPR) of 92.1\% and an F1 score of 94.6\%. The fine-tuned GPT-4o model outperforms other LLMs in distinguishing true positives from false positives. These results demonstrate that fine-tuning, even with a limited dataset, can substantially enhance detection performance.

\subsection{Large-Scale Measurement}
\label{sec:findings}

To understand the prevalence of \termname terms, we deploy the fine-tuned GPT-4o model with \platform for detection. The backend detection system was applied to English shopping websites filtered from the Tranco list's top 100,000 sites, along with two fake e-commerce website datasets: the FCWs dataset~\citep{bitaab2023beyond} and the FLOS dataset~\citep{janaviciute2023fraudulent}. This large-scale measurement serves as a qualitative study on the prevalence of \termname terms in popular shopping websites. We present our findings below. 

\label{sec:categoring}

\input{fig_tex/large_scale_measurement_stats}
\myparagraph{Categorizing Websites with \TermName Terms}
As shown earlier in~\autoref{table:dataset_stats},  we collect terms and conditions from \websitecnt English shopping websites, resulting in \termcnt terms. Using a GPT-4o model with the \termname term taxonomy, 10,150 terms (approximately \termpct) were flagged as \termname terms. Notably, \websitepct (3,471 out of 8,251) of the English shopping websites from the top 100,000 Tranco-ranked sites contain at least one type of non-legal \termname term. ~\autoref{fig:large_scale_stats}(a) and (b) show the number of terms and \termname terms across 8,251 websites, underscoring how difficult it is for consumers to review lengthy T\&Cs and pinpoint questionable financial terms thoroughly. This emphasizes the importance of automated detection systems to protect users from unfavorable terms.

\myparagraph{Trend Analysis}
\autoref{fig:large_scale_stats}(c) shows the distribution of unfavorable financial terms across categories in the top 100K Tranco-ranked websites~\citep{tranco}. Post-purchase terms (yellow) are the most common across all ranking levels, with a higher concentration in lower-ranked sites, suggesting these terms are more frequent on less popular websites. Purchase and billing terms (blue) also have significant representation. Termination and account recovery terms (red) and legal terms (green) are less frequent but more evenly spread across the rankings. This trend highlights the widespread presence of unfavorable financial terms, especially on lower-ranked sites, underscoring the need for greater regulation to protect consumers from harmful practices, particularly on less reputable websites.

\myparagraph{Comparing ShopTC-100K with Fake E-commerce Datasets}
Interestingly, the percentage of websites with \termname terms from the Tranco list (\websitepct) is similar to that of fraudulent e-commerce websites (46.70\%). This suggests that unfavorable financial terms are not limited to fraudulent sites but are also prevalent among high-ranking websites, pointing to a broader issue in consumer protection. \textit{ShopTC-100K} websites have more \termname legal terms, indicating that legitimate websites are more inclined to shift liability onto customers than fraudulent ones.

\myparagraph{Qualitative Study on User rating}
From the English shopping websites in the top 100k Tranco list, we select those with the highest frequency of \termname terms across categories. We analyze Trustpilot~\citep{trustpilot} reviews for the top 10 websites in each \termname term category with the highest presence, alongside 40 randomly selected websites. 
As shown in~\autoref{fig:large_scale_stats}(d), websites with \termname terms tend to have lower Trustpilot ratings, particularly those with ``Post-Purchase Terms'' and ``Purchase and Billing Terms,'' indicating negative customer satisfaction. ``Termination and Account Recovery'' and ``Legal Terms'' also correlated with lower ratings, though with more variation, suggesting mixed experiences. This suggests a link between \termname terms and consumer dissatisfaction.

\myparagraph{Qualitative Study on Current Ecosystem Defense}
We examine whether the top 10 websites with the highest frequency of \termname terms are flagged by ScamAdviser~\citep{scamadviser2024website}, Google Safe Browsing~\citep{google2024safebrowsing}, and Microsoft Defender SmartScreen~\citep{microsoft2024smartscreen}. Out of 40 websites, only 6 (15\%) have a ScamAdviser score below 90, and 5 (12.5\%) scored below 10, while the majority receive a perfect score of 100. None of the websites are flagged by Google Safe Browsing or Microsoft Defender, which is expected since \termname terms are not inherently indicative of scams. 

\myparagraph{Qualitative Study on User Perception} To illustrate the potential harm of \termname terms,  we present four case studies on user perception and financial harm in each category in Appendix~\ref{sec:case_studies}. This underscores the urgent need for automated systems to detect \termname terms effectively.

%% file: tables/classification_results.tex
\begin{table*}[!ht]
\footnotesize
\centering
\caption{Performance of LLMs in detecting \termname terms evaluated using zero-shot classification with GPT-3.5-Turbo, GPT-4-Turbo, GPT-4o, Llama 3 8B, and Gemma 2B, along with fine-tuned GPT-3.5-Turbo and GPT-4o models.}
\label{table:results}
\begin{tabular}{p{1.5cm} p{2cm} p{1.1cm} p{1.1cm} p{1.1cm} p{1.1cm} p{1.1cm} p{1.1cm} p{1.1cm} p{1.1cm}}
\toprule
\multirow{2}{*}{\textbf{Configuration}} & \multirow{2}{*}{\textbf{Model}} & \multicolumn{4}{c}{\textbf{Simple Prompt}} & \multicolumn{4}{c}{\textbf{Unfavorable Term Taxonomy Prompt}} \\
\cmidrule{3-10}
& & \textbf{FPR (\%) (\(\downarrow\) better)} & \textbf{TPR (\%)  (\(\uparrow\) better)} & \textbf{F1 (\%) (\(\uparrow\) better)} & \textbf{AUC (\%)  (\(\uparrow\) better)} & \textbf{FPR (\%)  (\(\downarrow\) better)} & \textbf{TPR (\%)  (\(\uparrow\) better)} & \textbf{F1 (\%)  (\(\uparrow\) better)} & \textbf{AUC (\%)  (\(\uparrow\) better)} \\

\midrule
\multirow{4}{*}{\parbox{4cm}{Zero-Shot}} 
& GPT-3.5-Turbo & 59.5 & 71.9 & 59.9 & 56.2 & 74.0 & 96.5 & 68.5 & 61.2\\
& GPT-4o & 58.6 & 72.6 & 61.4 & 56.2 & \textbf{34.4} & 96.6 & \textbf{82.5} & \textbf{80.3} \\
& GPT-4-Turbo & 58.6 & 72.6 & 61.4 & 56.2 & 64.8 & \textbf{100.0} & 73.8 & 67.2 \\
& Llama 3 & 73.5 & 82.3 & 61.4 & 54.4 & 48.5 & 86.7 & 71.3 & 69.1\\
& Gemma & 56.8 & 80.5 & 65.2 & 61.9 & 74.2 & 100.0 & 69.8 & 62.8\\
\midrule
{Fine-Tuned}
& GPT-3.5-Turbo & -  & -  & -  & - & 5.5 & 91.5 & 92.7 & \textbf{91.8} \\
& GPT-4o & -  & -  & -  & - & \textbf{2.3} & \textbf{92.1} & \textbf{94.6} & \textbf{91.8} \\

\bottomrule
\end{tabular}

\end{table*}

%% file: fig_tex/large_scale_measurement_stats.tex
\begin{figure*}[ht!]
\centering
\caption{Statistics from Large-scale measurement of \termname term detection on Tranco top 100K websites.}
\Description[Statistical analysis of unfavorable financial terms across shopping websites.]
 {This figure presents multiple statistical analyses of unfavorable financial terms detected on shopping websites from the Tranco top 100K list.
 (a) CDF of the number of terms per website, showing how frequently terms appear on different websites.
 (b) CDF of the number of unfavorable financial terms per website, illustrating the distribution of problematic terms across the dataset.
 (c) Distribution of unfavorable financial terms across different categories of websites based on their Tranco ranking, highlighting differences between highly ranked and lower-ranked sites.
 (d) Comparison of Trustpilot ratings between the top 10 websites with the most unfavorable financial terms and a random sample of 40 websites, analyzing whether websites with more unfavorable terms tend to have lower consumer ratings.}

\label{fig:large_scale_stats}
\subfigure[CDF of the number of terms per website.]{
\includegraphics[width=0.18\textwidth,height=0.18\textwidth]{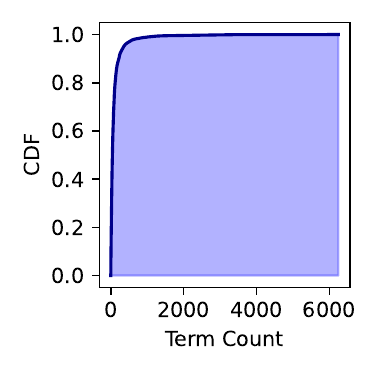}}
\subfigure[CDF of the number of unfavorable financial terms per website.]{
\label{fig:term_length_cdf}
\includegraphics[width=0.18\textwidth,height=0.18\textwidth]{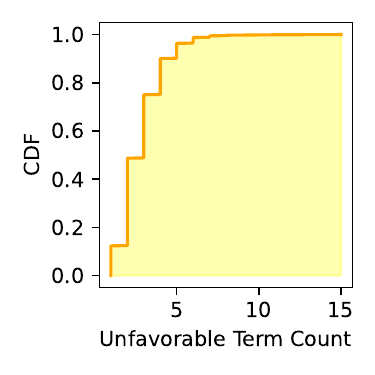}}
\subfigure[Distribution of unfavorable financial terms in each category across Tranco-ranked websites.]{
\label{fig:tranco_rank_dist}
\includegraphics[width=0.26\textwidth,height=0.18\textwidth]{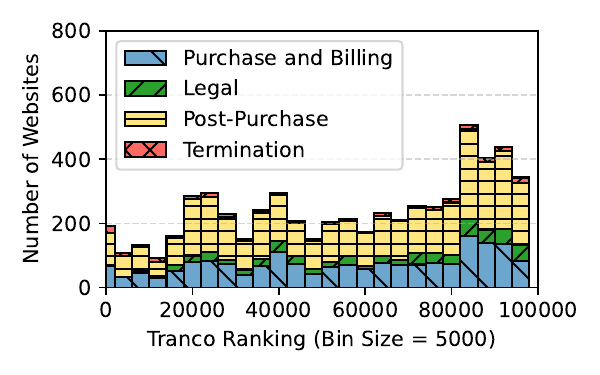}}
\subfigure[Trustpilot ratings comparison between the top 10 websites with the most unfavorable financial terms and a random sample of 40 websites.]{
\label{fig:imagnet-comparison}
\includegraphics[width=0.26\textwidth,height=0.18\textwidth]{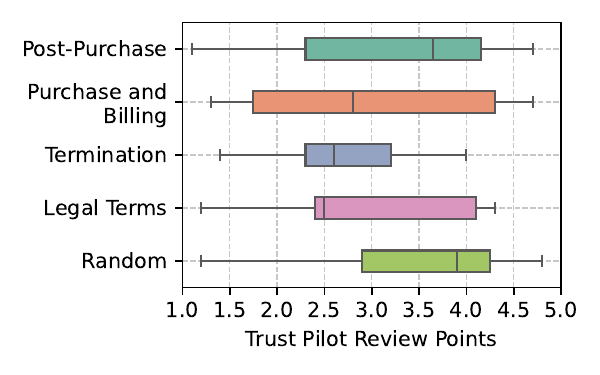}}

\end{figure*}

%% file: 06limitation.tex
\section{Discussion}
\label{sec:limitation}

We introduce \textit{TermMiner}, an open-source automated pipeline for collecting and modeling unfavorable financial terms in shopping websites with limited human involvement. Researchers can utilize our tools to examine various aspects of web-based text, such as readability or accessibility, and to conduct longitudinal studies.

\platform assumes that the financial terms in question are not \textit{adversarially perturbed}. Recent studies have highlighted LLM vulnerabilities to jailbreak and prompt injection attacks~\citep{zou2023universal, liu2023autodan, greshake2023not}. These attacks can result in incorrect outputs. However, for T\&Cs, such adversarial perturbations are likely to be subjected to manual scrutiny, particularly in post-complaint scenarios, such as legal disputes~\citep{celsius}. We leave the exploration of adversarial robustness in LLM-based \termname term detection for future work.

%% file: 07conclusion.tex
\section{Conclusion}

This paper is one of the first attempts to understand, categorize, and detect \termname terms and conditions on shopping websites. These terms, which can significantly impact consumer trust and satisfaction, have not been extensively studied. By highlighting the prevalence and types of \termname terms, we hope to pave the way for increased awareness and further research in this area. We develop an automated data collection and topic modeling pipeline, analyzing \termcnt terms from \websitecnt websites to create a taxonomy for \termname terms. This taxonomy includes \termtypecnt types across \termcatcnt categories, covering purchase and billing, post-purchase activities, account termination, and legal aspects.

\platform is the first study to evaluate the effectiveness of LLMs in identifying \termname terms. Using a fine-tuned GPT-4o model on a manually annotated dataset, \platform achieves an F1 score of 94.6\% with a false positive rate of 2.3\%. In large-scale deployment, we find that approximately \websitepct of shopping websites in the Tranco top 100,000 contain at least one category of \termname terms. Our qualitative analysis shows that current ecosystem defenses are inadequate to protect users from these terms, that less popular websites are more likely to include \termname terms, and that there is a correlation between \termname terms and user dissatisfaction.

\section*{Acknowledgments}

The authors would like to thank Renuka Kumar for her valuable input and suggestions during the early stages of this project.  This work is supported by a gift from the OpenAI Cybersecurity Grant program. Any opinions, findings, conclusions, or recommendations expressed in this material are those of the authors
and do not necessarily reflect the views of OpenAI.

%% file: 00main.bbl

%% file: appendix.tex
\appendix

\section*{Appendix}
\label{sec:appendix}

\section{Ethics}

We query and crawl terms and conditions from online shopping websites, collecting data from each site only once. Since terms and conditions are usually limited to a few subpages, this process does not overburden the servers hosting these websites. In this paper, we report some terms and conditions along with the associated companies, all of which are publicly available information. No personal data is collected during the measurement process.

\section{Financial Terms Topic Template}
\label{sec:appendix_finaincial_terms}

The financial term taxonomy consists of \financialcnt categories:

\begin{enumerate}
    \item \textit{Subscription/Product Terms}: Terms related to subscription fees, billing, and automatic renewals.

    \item \textit{Service Termination Policy}: Terms outlining the financial implications of service termination.
    \item \textit{Payment and Purchase Term}: Terms about payments, processing fees, and currency transactions.
    \item \textit{Return and Refund Policy}: Terms governing product returns and service cancellations.
    \item \textit{Insurance and Warranty Term}: Terms related to coverage, claims, limitations, and premiums.
    \item \textit{Promotions and Rewards}: Terms about offers, discounts, loyalty programs, and rewards.
    \item \textit{Shipping and Handling Terms}: Terms on product shipping costs and policies.
    \item \textit{Dispute Resolution Policy}: Terms outlining dispute resolution methods and processes.
    \item \textit{Investment and Trading Terms}: Terms specific to investment/trading platforms.
    \item \textit{Intellectual Property Terms}: Terms on rights and restrictions for intellectual property use.
    \item \textit{Financial Glossary}: Financial terminology definition.
    \item \textit{Others}: Includes less frequently mentioned financial terms.
\end{enumerate}

\section{Classification Prompts}
\label{sec:prompts}
Below is the simple binary prompt (``Simple Prompt'') used for zero-shot unfavorable financial term classification evaluation.
\begin{lstlisting}[label=Simple Prompt]
Classify the following term as 'malicious' or 'benign'. A term is 'malicious' if it is a financial term that is one-sided, unbalanced, unfair, or harmful to users. 

Respond only with 'm' for malicious or 'b' for benign.
\end{lstlisting}

Below is the multi-class taxonomy prompt (``Unfavorable Term Taxonomy Prompt'') used for zero-shot evaluation of unfavorable financial term classification and for fine-tuning LLMs.
\begin{lstlisting}[label=Unfavorable Term Taxonomy Prompt]
You will be provided with a paragraph extracted from the terms and conditions. Your task is to classify them into one of the topics below or 'b' for 'benign':

- Automatic Subscription after Free Trial: Automatically subscribing users after free trials.
- [...]

If the term is not a financial term or a reasonable financial term based on common sense, reply 'b' for 'benign'.

If the term is malicious and financial, reply with a topic from the template above. 
\end{lstlisting}

\section{Shopping website classification Evaluation}
\label{sec:website_cls}

\input{tables/website_cls}

To evaluate our classification methods, we randomly sampled 500 websites from the top 10,000 on the Tranco list for manual annotation to identify shopping sites (offering products or services for sale). Of these, 257 were categorized as ``shopping'', 219 as ``non-shopping'', and 24 were inaccessible, due to either server issues or geo-blocking IPs in the US.
As indicated in~\autoref{table:website_cls}, GPT-4o's accuracy reached 86\% when analyzing URLs with corresponding screenshots. Further examination of a focused group, specifically English-language websites with available T\&Cs (115 out of 500), reveals that accuracy improved to 81\% for GPT-3 using only URLs and to 92\% for GPT-4o with screenshots, approaching the commercial-grade classification benchmarks reported in previous studies (89\%-93\%)~\citep{mathur2019dark}. The FPR of the focused group is 6.3\%. Based on these statistics, GPT-4o paired with image data is selected for the broader scale measurement of shopping websites.

\section{\TermName Terms Taxonomy}
\label{sec:detailed_tax}

We discover \termname term types falling under 4 broader categories: 1) purchase and billing terms; 2) post-purchase terms; 3) termination and account recovery terms; and 4) legal terms. We describe the taxonomy with an explanation for each type below:

\subsubsection{Unfavorable Purchase and Billing Terms}

This category includes subscription, purchase, and billing terms that are unfavorable or concerning for customers:

\begin{itemize}
    \item \textbf{Immediate Automatic Subscription.} 
    Additional subscriptions are automatically added when purchasing an item or during promotions without clear consent from the user.
    \item \textbf{Automatic Subscription after Free Trial.} Users are automatically enrolled in a paid subscription after a trial period ends unless they actively cancel.
    \item \textbf{Unilateral Unauthorized Account Upgrades.} Accounts may be upgraded and charged at higher rates without providing prior notice to the user.
    \item \textbf{Late or Unsuccessful Payment Penalty.} Penalties or interest charges are applied for late or unsuccessful payments.
    \item \textbf{Overuse Penalty.} Charging extra fees if usage limits are exceeded. Typically found in subscription-based services such as data plans, cloud storage, and streaming services.
    \item \textbf{Retroactive Application of Price Change.} 
    Price (of sub-scription-based services) increases can be applied retroactively without prior notice to the user.
\end{itemize}

\subsubsection{Unfavorable Post-Purchase Terms }
This category includes cancellation, shipping, return, and refund terms that are unfavorable or concerning for customers:

\begin{itemize}
    \item \textbf{Non-Refundable Subscription Fee.} Subscription fees that have already been charged are not refunded.
    \item \textbf{No Refund for Purchase.} Purchases of individual items are final and non-refundable.
    \item \textbf{Strict No Cancellation Policy.} Orders cannot be canceled once they have been processed.
    \item \textbf{Cancellation Fee or Penalty.} Fees are applied for canceling certain bookings, services, or online purchase orders.
    \item \textbf{Non-Refundable Additional Fee.} Charging non-refundable additional fees under various labels such as service fees, transfer fees, pre-authorization fees, administrative fees, subscription upgrade fees, handling product fees, etc.
    \item \textbf{Non-Monetary Refund Alternatives.} Refunds are provided in the form of coupons, reward points, or store credit rather than money.
    \item \textbf{No Responsibility for Delivery Delays.} Companies are not held liable for delays in product delivery.
    \item \textbf{Customers Responsible for Shipping Issues.} Customers are responsible for handling customs issues, additional shipping charges, and any shipping-related complications that do not involve delays.
    \item \textbf{Customers Pay Return Shipping.} Customers bear the cost of return shipping for products.
    \item \textbf{Restocking Fee.} A fee is charged for restocking returned items.
\end{itemize}

\subsubsection{Unfavorable Termination and Account Recovery Terms}
This category includes account or service termination, deactivation, and reactivation terms that are unfavorable or concerning for customers:

\begin{itemize}
    \item \textbf{Account Recovery Fee.} A Fee is charged to recover locked or archived accounts.
    \item \textbf{Digital Currency, Reward, Money Seizure on Inactivity.} Digital assets, such as rewards, points, and virtual currencies, are forfeited or otherwise taken away if accounts remain inactive for extended periods.
    \item \textbf{Digital Currency, Reward, Money Seizure on Termination or Account Closure.} Digital assets, such as rewards, points, and virtual currencies, are forfeited or otherwise taken away upon service termination or account closure.
\end{itemize}

\subsubsection{Unfavorable Legal Terms}
This category includes legal terms that are unfavorable or concerning for customers:
\begin{itemize}
    \item \textbf{Exorbitant Legal Document Request Fee.} High fees are charged for requesting legal documents.
    \item \textbf{Forced Waiver of Legal Protections.} Customers are required to waive certain legal protections.
    \item \textbf{Forced Waiver of Class Action Rights.} Customers waive their rights to participate in class action lawsuits.
    \item \textbf{Other Legal Unfavorable Financial Term.} Additional legal terms that impose financial burdens or limit legal recourse for the customer.
\end{itemize}

\input{fig_tex/scam_example2}
\input{fig_tex/neteller_review}

Many terms and conditions for online shopping websites include strong legal language, such as waivers of class action rights, arbitration clauses, and limitations of liability. This study does not specifically focus on the legal aspects for two main reasons: (1) Although legal terms can impact users financially, they differ from other categories we report. These terms, despite their potential future implications, are not the primary concern when customers make a purchase or sign up for a service. (2) There is another line of work (see Appendix~\ref{sec:appendix_other_templates}) that focuses on terms deemed invalid in court. We consider our work complementary to these studies. By not intensively focusing on legal terms, we maintain a focus on terms with immediate and direct financial implications for users.

\section{Case Studies}
\label{sec:case_studies}
We present four case studies illustrating the potential harm of \termname terms in each \termname term category.

\myparagraph{Unfavorable post-purchase terms case study} National Park Reservations~\citep{nationalparkreservations}, a company providing national park hotel and lodging reservation service with a 1-star review on Yelp, includes the following terms and conditions:

\begin{quote}
    For this service, National Park Reservations charges a 10\% non-refundable reservation fee based on the total dollar amount of reservations made. This reservation fee will be billed separately to your credit card and will be billed under the memo “National Park Reservations”. By using National Park Reservations, the customer authorizes National Park Reservations to charge their credit card the 10\% non-refundable fee.
\end{quote}

\input{fig_tex/wordcloud}

The above terms fall under the ``Non-Refundable Additional Fee'' category. \autoref{fig:wordcloud} shows a word cloud that displays the most frequent words from the top 50 Yelp reviews (2021-2024), excluding the company name. Despite the non-refundable additional fee being clearly stated in the terms and conditions, many customers still find it deceptive. ``Scam'' is among the most frequent words in the reviews. This shows the potential harm caused by \termname terms and perceived deceptive practices, significantly impacting customer trust and satisfaction.

\myparagraph{Unfavorable termination and account recovery terms case study} 
Compared to other categories of \termname terms, those related to unfair, unfavorable, or concerning service termination and account management are significantly less prevalent in our measurements, as shown in~\autoref{fig:large_scale_stats}. However, these terms can still impose substantial costs on customers. For example, the Terms of Use from Neteller~\citep{neteller}---a digital wallet with a TrustScore of 10 out of 100 from Scamadviser~\citep{scamadviser2019algorithm}---include such terms:

\begin{quote}
    If an Account has been closed, [...]. Fees relating to ongoing management of inactive accounts will also continue to be charged following the closure of your Account. This provision shall survive the termination of the relationship between you and us.
\end{quote}

This term means that even after an account is closed, fees for managing inactive accounts will continue to be charged, potentially resulting in unexpected costs for customers. See~\autoref{fig:neteller} in the Appendix for a compilation of customer complaints about account closures and issues with retrieving deposited funds. Similar to~\autoref{fig:example2}, the terms from Neteller demonstrate the potential harm of unfavorable termination and account recovery terms.

\myparagraph{Unfavorable legal terms case study} Another concerning \termname term is the inclusion of a waiver for California Civil Code Section 1542, found in 2.1\% (152 out of 7,225) of the top 80,000 Tranco-ranked websites. An example of such a term states:

\begin{quote}
If you are a California resident, you shall and hereby do waive California Civil Code Section 1542, which says: “A general release does not extend to claims which the creditor does not know or suspect to exist in his favor at the time of executing the release, which, if known by him must have materially affected his settlement with the debtor.”
\end{quote}

California Civil Code Section 1542 protects individuals from unknowingly giving up their right to make claims for issues they were not aware of at the time they signed a release. By including a waiver for this code, websites are essentially asking customers to give up this important protection. This indicates that these sites are aware that most people do not thoroughly read the terms and conditions. This oversight can be leveraged to disable significant legal protection, which can make co-existing \termname terms harder to dispute. In fact, 60.5\% (92 out of 152) of the websites with the California Civil Code Section 1542 waiver also have at least one other category of \termname terms.

\section{Comparison with Other Online Agreement Annotation Scheme}
\label{sec:appendix_other_templates}

In this section, we introduce the annotation templates proposed under the European Union (EU) framework for identifying unfair contract terms~\citep{loos2016wanted, galassi2024unfair, galassi2020cross, lippi2019claudette, drazewski2021corpus}. While these studies emphasize legal categories and jurisdictional issues, our research specifically targets financial terms in online service agreements. 

Loos \etal~\citep{loos2016wanted} analyze the unfair contract terms of online service providers in light of the Unfair Contract Terms Directive (UCTD)~\citep{CouncilDirective1993} of the European Union. The authors examine various types of contractual terms from international online service providers, identifying those that are unlikely to pass the Directive’s fairness test in the following five categories:

\begin{itemize}
    \item \textbf{Unilateral Changes}: Our analysis also considers unilateral changes made by online service providers, particularly regarding financial aspects such as unilateral price changes, plan upgrades, and various penalties.
    \item \textbf{Termination Clauses}: We examine termination clauses focusing on their financial consequences, including the seizure of digital currency, reward points, or money upon termination.
    \item \textbf{Liability Exclusions and Limitations}: Both our paper and the authors' findings highlight the problematic nature of liability exclusions and limitations, which often unjustly limit the providers' responsibility for service failures, thereby creating a significant imbalance in the parties' rights and obligations.
    \item \textbf{International Jurisdiction and Choice-of-Law Clauses}: Although we have a category for unfavorable legal terms, the detailed categorization of unfair legal terms is deferred to future work. This is because, compared to other unfavorable financial terms, legal terms typically have a more indirect impact on users.
    
    \item \textbf{Transparency}: While the readability and accessibility of terms and conditions are not the main focus of this paper, our data collection and topic modeling pipeline can be readily adapted for future research in these areas.
\end{itemize}

Another worth-noting line of work in unfair online agreements~\citep{galassi2024unfair, galassi2020cross, lippi2019claudette, drazewski2021corpus}. The CLAUDETTE~\citep{lippi2019claudette} system evaluates the fairness of terms within the jurisdiction of the European Union by leveraging legal standards and principles established within the EU framework. The annotation scheme is as follows:

\begin{itemize}
    \item Jurisdiction for disputes in a country different from the consumer’s residence.
    \item Choice of a foreign law governing the contract.
    \item Limitation of liability.
    \item The provider’s right to unilaterally terminate the contract/access to the service.
    \item The provider’s right to unilaterally modify the contract/service.
    \item Requiring a consumer to undertake arbitration before court proceedings can commence.
    \item The provider retaining the right to unilaterally remove consumer content from the service, including in-app purchases.
    \item Having a consumer accept the agreement simply by using the service, without having to click on “I agree/I accept”
    \item The scope of consent granted to the ToS also takes in the privacy policy, forming part of the “General Agreement”
\end{itemize}

We consider our work to be a complementary addition to the AI \& Law database, with our template being more aligned with the natural phrasing found in terms and conditions of online shopping websites. We hope that future research will incorporate both the legal templates and our proposed template to provide a more comprehensive understanding of the landscape of unfair (financial) terms.

\section{Term Page Regex}
\label{sec:appendix_reg}

Below are the positive and negative regex matching pattern we deploy for this work:

\begin{lstlisting}[label=Positive]
positive_regex = [
    "terms.*?conditions",
    "terms.*?of.*?use",
    "terms.*?of.*?service",
    "terms.*?of.*?sale",
    "terms.*?of.*?conditions",
    "terms.*?and.*?conditions",
    "terms.*?&.*?conditions",
    "conditions.*?of.*?use",
    "intellectual.*property.*policy",
    "return[s]?.*?policy",
    "refund[s]?.*?policy",
    "return.*?and.*?refund.*?policy",
    "cancellation.*?and.*?returns",
    "cancellation.*?returns",
    "prohibited.*conduct",
    "electronic.*communication.*policy",
    "safety.*guideline",
    "requests.*from.*law.*enforcement",
    "bonus.*terms.*apply",
    "community.*rules",
    "gift.*card.*policy",
    "contact.*us.*here",
    "shipping.*policy",
    "warranty",
    "end.*user.*license",
    "user.*?agreement",
    "payment.*terms",
    "content.*policy",
    "terms"
]

\end{lstlisting}

The negative regex list is as follows:
\begin{lstlisting}[label=Negative]
negative_regex = [
    "privacy.*?policy",
    "cookie.*?policy",
    "privacy.*?notice",
    "sale.*?tax.*?policy",
    "prohibited.*?items",
    "1099.*?k.*?form",
    "dmca.*copyright.*notification",
]

\end{lstlisting}

%% file: tables/website_cls.tex
\begin{table}[t!]
\centering
\caption{Performance and Cost Analysis for Zero-Shot Shopping Website Classification Using GPT-3.5-Turbo~\citep{gpt35} and GPT-4o~\citep{openai2023gpt4} Models on 500 randomly selected websites.}
\footnotesize
\begin{tabular}{p{15mm} p{13mm} p{4mm} p{4.5mm} p{4.5mm} p{4.5mm} p{4mm} }
\toprule
{Model} & {Configuration} & {ACC (\%)} & {TP (\%)} & {FP (\%)} & {TN (\%)} & {FN (\%)}\\ 
\midrule
\multirow{2}{*}{{GPT-3.5-Turbo}} &
URL & 73 & 80.9 & 36.5 & 63.5 & 19.1  \\
& URL+HTML & 24 & 31.5 & 5.5 & 94.5 & 68.5 \\
\midrule

\multirow{2}{*}{{GPT-4o}} &
URL & 63 & 40.9 & 10.1 & 89.9 & 59.1 \\
& URL+Image & 86 & 90.7 & 20.6 & 79.4 & 9.3  \\

\bottomrule
\end{tabular}

\label{table:website_cls}
\end{table}



%% file: fig_tex/scam_example2.tex
\begin{figure}[!t] 
 \centering
 \includegraphics[width=0.99\columnwidth]{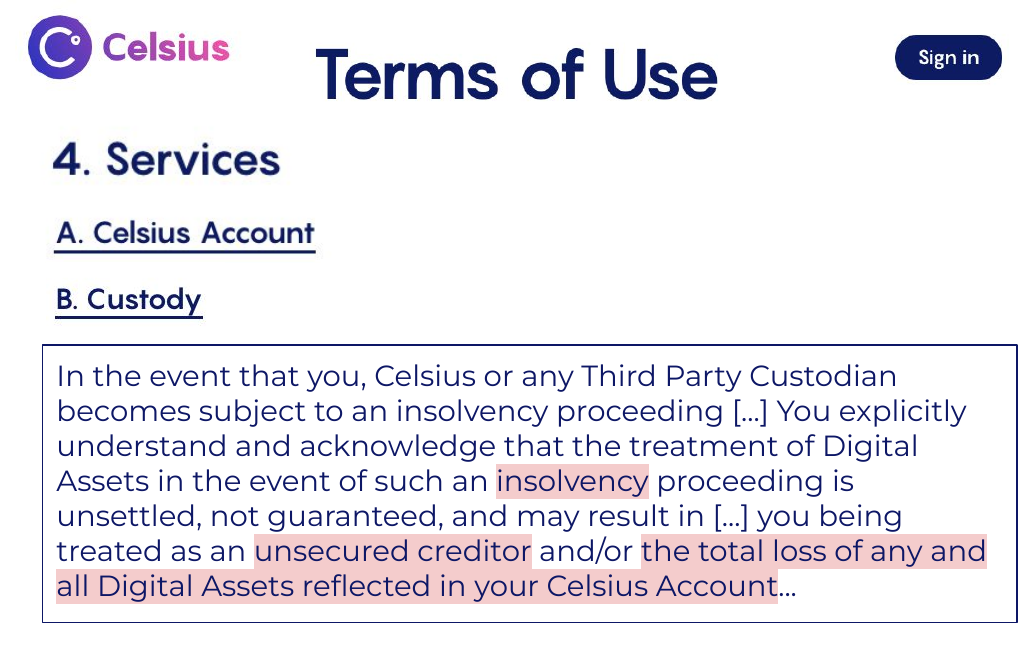}
 \caption{Extracted from the T\&C of Celsius Network LLC, a now bankrupt cryptocurrency company. }
 \Description[Excerpt from Celsius Network LLC’s terms and conditions.]
 {This figure presents a term extracted from the terms and conditions (T\&C) of Celsius Network LLC, a cryptocurrency company that later filed for bankruptcy. 
 The excerpt highlights specific clauses related to user funds and liabilities, illustrating how unfavorable financial terms may have impacted customers.}
\label{fig:example2}
\end{figure}

%% file: fig_tex/neteller_review.tex
\begin{figure}[t] 
 \centering
 \includegraphics[width=0.97\columnwidth]{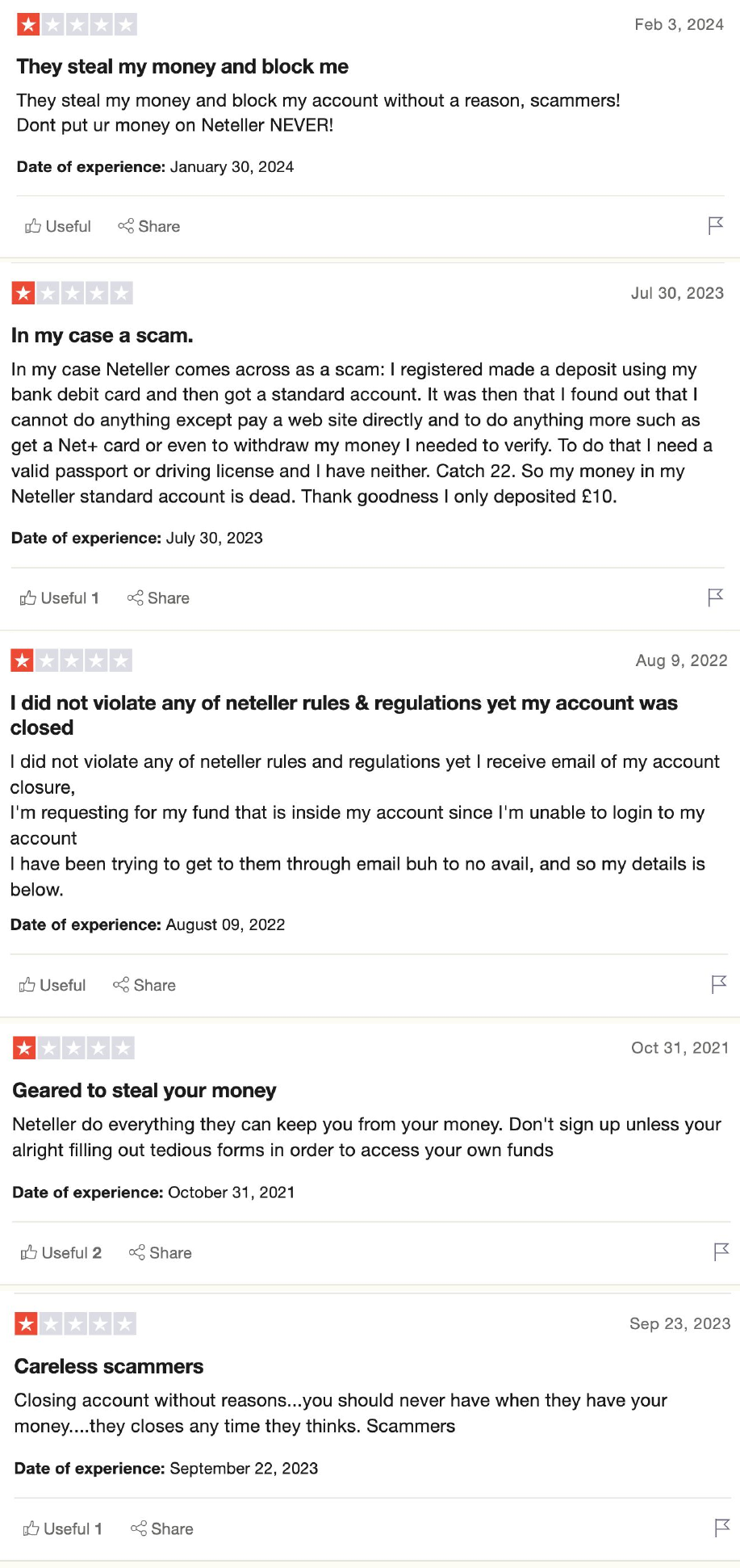}
 \caption{Complied reviews for Neteller~\citep{neteller} from Trust Pilot~\citep{trustpilot} regarding account closure.}
 \Description[Trustpilot reviews of Neteller account closures.]
 {This figure presents compiled user reviews from Trustpilot regarding account closures by Neteller.
 The reviews highlight customer complaints about sudden account terminations, withheld funds, and difficulties in recovering their balances.
 The feedback provides insight into customer dissatisfaction with Neteller’s account closure policies.}
\label{fig:neteller}
\end{figure}

%% file: fig_tex/wordcloud.tex
\begin{figure}[!t] 
 \centering
 \includegraphics[width=0.99\columnwidth]{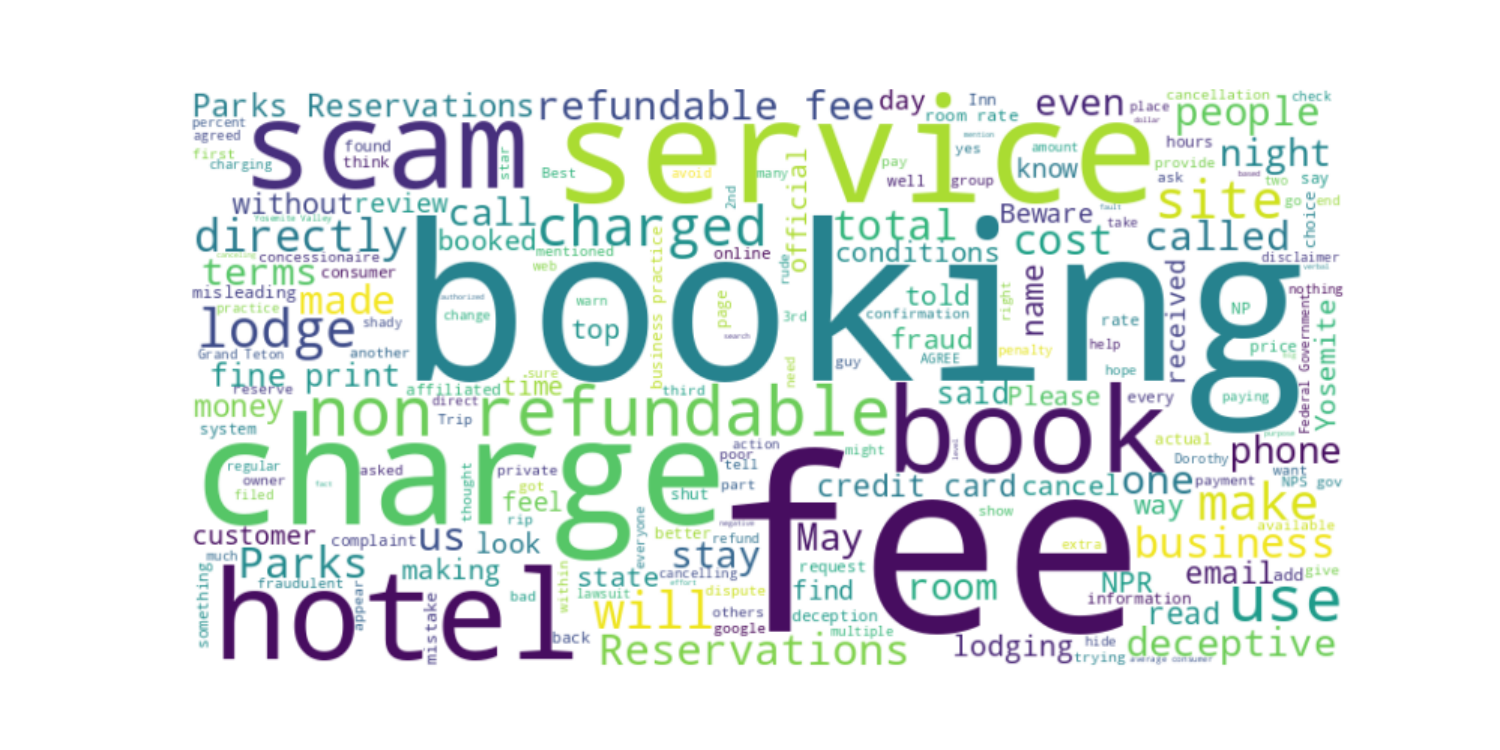}
 \caption{Word cloud based on the top 50 Yelp reviews of National Park Reservations, whose T\&Cs specify a 10\% non-refundable booking fee. ``Scam'' and ``non-refundable'' are frequently mentioned words in the reviews.}
 \Description[Word cloud of Yelp reviews for National Park Reservations.]
 {This figure displays a word cloud generated from the top 50 Yelp reviews of National Park Reservations.
 The company's terms and conditions specify a 10\% non-refundable booking fee, which is a common theme in customer complaints.
 Frequently appearing words in the reviews include ``scam'' and ``non-refundable,'' indicating recurring concerns from users regarding the booking policy.}
\label{fig:wordcloud}
\end{figure}